\documentclass[runningheads]{llncs}
\usepackage[T1]{fontenc}

\usepackage{graphicx}
\usepackage{amsmath}

\usepackage{algorithm}
\usepackage{algpseudocode}
\usepackage{dblfloatfix}
\usepackage{booktabs}
\usepackage{tabularx}

\usepackage{array}
\usepackage{makecell}
\usepackage{float}
\usepackage[most]{tcolorbox}
\usepackage{multicol}
\usepackage{enumitem}
\usepackage{amssymb}
\usepackage{placeins}
\usepackage{hyperref}
\usepackage{tikz}
\usetikzlibrary{positioning, arrows.meta}

\begin{document}
\title{CEDAR-42001: From ISO/IEC 42001 Conformity to Architecture-Aware, Audit-Visible Assurance Posture for AI Cyber-Physical Systems}
\titlerunning{CEDAR-42001: Conformity to Posture for AI-CPS}

\author{Priyanka Prakash Surve\inst{1}\orcidID{0009-0001-5673-2687} \and
Asaf Shabtai\inst{1}\orcidID{0000-0003-0630-4059} \and
Yuval Elovici\inst{1}\orcidID{0000-0002-9641-128X}}
\authorrunning{P. Surve et al.}

\institute{Ben Gurion University of the Negev \\
\email{surve@post.bgu.ac.il}\\
\email{shabtaia@bgu.ac.il}
\email{elovici@bgu.ac.il}}
\maketitle              % typeset the header of the contribution
\begin{abstract}
AI-enabled cyber-physical systems (AI-CPS) turn data-driven decisions into physical actions, creating assurance challenges across sensing, computation, control, human oversight, and governance. 
ISO/IEC 42001:2023 specifies requirements for an artificial intelligence management system (AIMS), but conformity assessment alone does not show which architectural layers are affected, whether practices are mature enough for the risk context, or what actions should follow.
We present CEDAR-42001 \footnote{Artifact and reproducibility materials: 
\url{https://doi.org/10.5281/zenodo.20722090}} (Control–Evidence Decision and Action Reasoning), a two-stage method that converts ISO/IEC 42001 audit evidence into an architecture-aware assurance posture traceable to the audit record. 
Stage A preserves the conformity determination. Stage B adds four outputs to each audit row: (i) attribution to a governance stratum or one of seven AI-CPS layers; (ii) a five-dimensional maturity profile with binding-constraint identification; (iii) a risk-proportionate target maturity; and (iv) a rulebook-derived action recommendation. 
The enriched rows are aggregated into strategic, operational, and tactical decision products.
We evaluate CEDAR-42001 using a synthetic autonomous-fleet AIMS and by comparing conformity-only results with the enriched outputs. 
Although 89.9\% of audit rows were conforming, only 34.3\% of conforming rows reached the baseline High-assurance category; across alternative operationalizations, this proportion ranged from 22.4\% to 46.2\%. 
A retrospective application to the 2023 Cruise robotaxi incident shows how the method captures documented concerns across governance, perception, decision-making, and human oversight and maps them to layer-specific actions. 
CEDAR-42001 does not estimate exploitability or replace technical CPS-security testing; it identifies where audit evidence warrants deeper technical assurance, organizational improvement, or remediation.

\end{abstract}

\section{Introduction}
\label{sec:introduction}

AI-enabled cyber-physical systems (AI-CPS), such as autonomous vehicles, mobile and humanoid robots, clinical systems, and AI-enabled industrial control systems, use data and AI-based decisions to control or influence physical processes, operational environments, and human safety. Ensuring assurance in these systems requires a cross-layer perspective because a weakness in one component can propagate through the system and affect other components or functions.
For example, a vulnerability exploited in sensing or perception can distort state estimation or decision-making, propagate through control and actuation, and produce consequences shaped by human supervision and organizational response.
Documented cases include LiDAR-induced trajectory-prediction errors, physical adversarial objects that cause system-level perception failures, and autonomous-system incidents involving technical, human, and governance breakdowns~\cite{lou2024first,wang2023does}.

The October 2023 Cruise robotaxi incident illustrates this interaction. 
After a pedestrian was propelled into the vehicle's path by another vehicle, the Cruise vehicle stopped but then initiated a programmed post-collision pullover while the pedestrian remained beneath it. 
The available record indicates failures across multiple system components: the vehicle did not maintain an adequate representation of the pedestrian's position; its decision and control logic selected an unsafe response to an unanticipated condition; and remote assistance could not intervene before the maneuver was completed. 
Investigations also identified deficiencies in organizational communication and incident reporting~\cite{quinnemanuel2024,koopman2024,cpuc2023osc,dotoig2024}.
The incident, therefore, involved sensing and perception, decision-making and control, physical action, human oversight, and governance.
It shows that AI-CPS assurance cannot be reduced to either technical failure analysis or management-system compliance.

ISO/IEC~42001:2023 specifies requirements for establishing, implementing, maintaining, and continually improving an artificial intelligence management system (AIMS)~\cite{iso42001}. 
These requirements address organizational context, risk management, operational control, performance evaluation, and continual improvement. 
Conformity assessment determines whether an AIMS satisfies these requirements, but does not validate the technical security or safety properties of the AI-CPS it governs.
Because the standard is organized around management-system functions rather than CPS architectural layers, a conformity verdict or list of nonconformities does not identify which system layers are implicated, whether relevant practices are sufficiently mature for their risk context, or which layer-specific actions should follow.

We define the bounded diagnosis derived from such governance evidence as the \emph{audit-visible assurance posture}.
For each assessed obligation, it records the relevant governance or AI-CPS layer, the maturity and binding constraint of the governed practice, the maturity required by its risk context, and the resulting action recommendation. 
This construct characterizes assurance supported by the assessed AIMS evidence; it does not measure the deployed system's full technical security posture, exploitability, vulnerability exposure, or runtime behavior.

This limitation creates a decision gap. 
Strategically, operators must determine how to allocate limited assurance resources. 
Operationally, they must identify priority weaknesses and quantify the gap between current and risk-required capabilities.
Tactically, assessors and control owners must select corrective or improvement actions. 
Clause-based conformity provides an essential governance baseline but does not directly answer these architecture- and action-oriented questions. 
Operational-technology practice similarly separates management-system governance from architecture-aware security assessment, as reflected in the complementary roles of ISO/IEC~27001 and the IEC~62443 series~\cite{iso27001,iec62443}.

We introduce CEDAR-42001 (Control-Evidence, Decision, and Action Reasoning), a two-stage method for this translation.
Stage~A preserves the original ISO/IEC~42001 conformity determination. 
Stage~B augments each audit row with four diagnostic fields: attribution to a governance layer or one of seven AI-CPS architectural layers; a five-dimensional maturity profile and its binding constraint; a risk-proportionate target maturity; and a rulebook-derived action recommendation. 
The enriched rows are then aggregated into strategic, operational, and tactical decision products.

CEDAR-42001 establishes a governance-to-assurance handoff by locating each governed practice or deficiency within the AI-CPS architecture and identifying the technical or organizational assurance activity supported by the available evidence.
Such activities may include penetration testing, architecture review, controller validation, runtime monitoring, safety analysis, or human-oversight evaluation. 
Section~\ref{sec:discussion} distinguishes this scope from technical CPS security assessment.

This paper makes two contributions. 
First, it formulates the decision gap between ISO/IEC~42001 conformity and the architecture-aware assurance information required by AI-CPS operators, and defines the resulting bounded construct as the audit-visible assurance posture. 
Second, it introduces CEDAR-42001, a method that retains the original conformity determination while enriching it with architectural attribution, maturity, and binding-constraint analysis, risk-proportionate target levels, and tailored action recommendations.
To our knowledge, no prior method co-locates conformity, CPS-layer attribution, maturity sufficiency, and action class within a single traceable audit row. 
Its principal methodological contribution is row-level traceability: every aggregate can be decomposed into an individual obligation, evidence note, and triggered rule.

We demonstrate the method through an end-to-end pipeline-coverage study, a comparison of conformity-only and enriched decision outputs, and a retrospective application to the documented Cruise robotaxi incident.
\section{Background and Related Work}
\label{sec:background}

CEDAR-42001 lies at the intersection of AI management-system conformity, capability assessment, and CPS security assurance.
Existing approaches largely treat these areas separately.

\subsection{AI Management-System Conformity}

ISO/IEC~42001:2023 specifies requirements for establishing, implementing, maintaining, and continually improving an artificial intelligence management system (AIMS) \cite{iso42001}. 
Clauses~4--10 cover organizational context, leadership, planning, support, operation, performance evaluation, and improvement, while Annex~A provides a reference set of AI controls.
Conformity assessment determines whether an organization meets the applicable requirements of its management system.

This function differs from evaluating the governed AI system itself.
ISO/IEC 42001 neither prescribes a CPS architecture nor assigns technical security levels to system components, and it does not certify the safety or security of a deployed AI-CPS. 
Its primary outcome is a conformity determination supported by audit evidence and documented nonconformities.
The same clause-level finding may therefore reflect weaknesses in different architectural layers, levels of operational maturity, or remediation needs.

CEDAR-42001 preserves this conformity function.
It begins with an audit row derived from an ISO/IEC~42001 obligation and its supporting evidence, then adds a separate assurance diagnosis without altering the original conformity determination.

\subsection{Maturity and Risk-Proportionate Assurance}

Capability maturity models in the SW-CMM lineage distinguish documented processes from repeatable execution, monitoring, measurement, and continual improvement~\cite{ISACA-CMMI}.
This distinction is important in AI-CPS governance because a formally established control may provide limited assurance if it is inconsistently applied, poorly monitored, or disconnected from the technical layers it governs.

Risk-management frameworks apply a related principle of proportionality. 
The NIST AI Risk Management Framework organizes risk-management activities across governance, mapping, measurement, and management~\cite{NISTAIRMF}. 
In operational technology, IEC~62443 and NIST SP~800-82 relate security expectations to system criticality, potential consequences, and operational context~\cite{iec62443,nist80082}. 
Together, these frameworks support the principle that the strength of a governed practice should be proportionate to the consequences of failure.

CEDAR-42001 adopts these principles without importing an external control set. 
It evaluates each governed practice in terms of traceability, operationalization, monitoring, improvement, and cross-layer integration, then compares the resulting maturity level with a target derived from CPS impact and safety or security criticality. 
This yields a row-level judgment of whether the evidence supporting an ISO/IEC~42001 obligation is sufficiently mature for its risk context.

\subsection{CPS Security, Auditing, and Assurance Evidence}

CPS and operational-technology security distinguish management governance from technical and architectural assurance. 
ISO/IEC~27001 addresses information-security management, whereas IEC~62443 addresses system architecture, zones and conduits, component requirements, and security levels~\cite{iso27001,iec62443}. 

Research on AI-enabled CPS examines technical dependability, platform exposure, and runtime assurance. 
RISK-MAP provides a quantitative cybersecurity assessment for layered robotic and humanoid systems, while DURA-CPS coordinates dependability-assurance activities for LLM-enabled CPS~\cite{surve2025sok,srinivasan2025dura}.
These approaches rely on system-, component-, implementation-, or runtime-level evidence rather than management-system audit records.
They therefore complement, rather than replace, governance assessment.

Algorithmic-auditing frameworks emphasize internal accountability, evidence collection, documentation, and independent review~\cite{raji2020smactr,lam2024assurance,mokander2023auditing}. 
Assurance-case research similarly structures claims, evidence, and arguments about system properties~\cite{bloomfield2025assurance}. 
Although these approaches demonstrate the value of traceable assurance artifacts, they do not integrate ISO/IEC 42001 conformity, AI-CPS layer attribution, maturity sufficiency, and recommended actions within a single audit row.

CEDAR-42001 provides this cross-domain handoff. 
It links an audit-visible practice or deficiency to its architectural location, evaluates whether its maturity is proportionate to the associated risk, and identifies the technical or organizational assurance activity that should follow.
The next section defines the assurance scope and the cross-layer failure model that supports this attribution.
\section{Assurance Scope and Cross-Layer Failure Model}
\label{sec:assurance-scope}

CEDAR-42001 applies to AI-enabled cyber-physical systems in which data-driven decisions may affect physical processes, operational environments, or human safety.
The governed system is organized into seven AI-CPS architectural layers adapted from Surve et al.~\cite{surve2025sok}: physical processes and actuation (L1), sensing and perception (L2), data processing and state estimation (L3), middleware and communication (L4), AI decision-making and control (L5), application logic and task execution (L6), and human supervision and interaction (L7). 
We also define a governance layer (G) for obligations and deficiencies related to the AIMS as a whole, rather than a single technical layer. 
Findings that cannot be meaningfully localized are classified as Cross-layer. This is a reporting category, not an additional architectural layer; portfolio-level summaries retain one primary attribution to avoid double-counting.

The assurance scope covers three classes of adverse events. 
\emph{Adversarial events} result from intentional manipulation, including sensor spoofing, adversarial inputs, unauthorized command injection, and compromise of communication or middleware. 
\emph{Technical and safety failures} include perception errors, incorrect state estimation, unsafe control logic, inadequate fail-safe behavior, and failure under unanticipated operating conditions.
\emph{Organizational and oversight failures} include ineffective monitoring, delayed human intervention, incomplete incident response, and deficiencies in internal or external communication. 
These classes may overlap: one incident may combine adversarial influence, technical failure, inadequate human supervision, and governance weaknesses.

The layered model captures how such concerns propagate through the system. 
Manipulated or erroneous sensor data at L2 may distort state estimation at L3, alter control decisions at L5, and produce physical consequences at L1.
The severity of those consequences may depend on whether L7 supports timely intervention and whether G provides adequate monitoring, escalation, review, and corrective action.
Thus, the layer at which harm becomes observable may differ from the layer in which the governed weakness originates.

CEDAR-42001 uses this model as the attribution vocabulary for Stage~B.
For each assessed obligation, the assessor identifies the governance layer or AI-CPS layer implicated by the evidence.
This attribution is combined with maturity, risk context, and the binding capability constraint to determine the required assurance or remediation activity.

\begin{figure}[t]
\centering
\includegraphics[width=0.84\textwidth]
{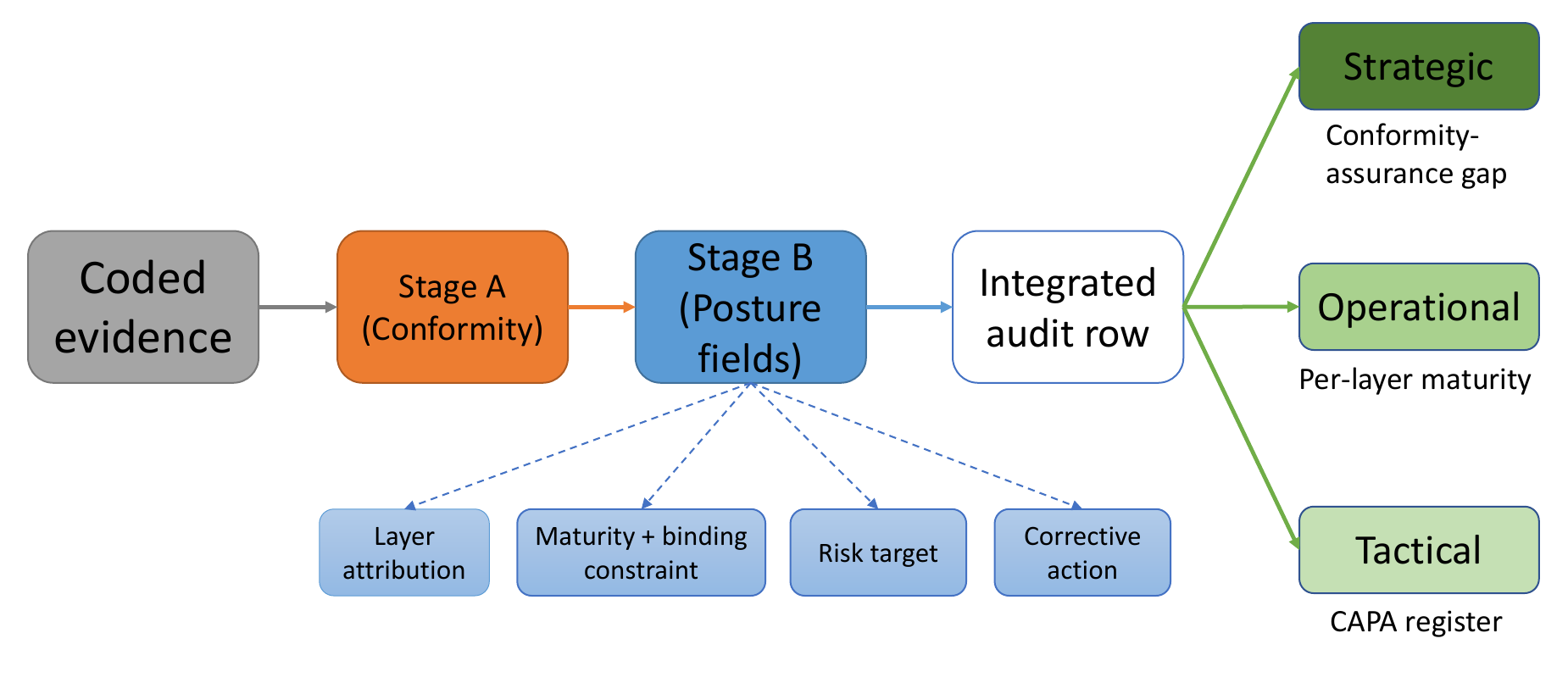}
\caption{Two-stage CEDAR-42001 row-level workflow. Stage~A determines conformity; Stage~B adds the four posture fields. Integrated rows feed strategic, operational, and tactical decision products.}
\label{fig:cedar-overview}
\end{figure}

Attribution, therefore, identifies where a governed weakness resides and how its effects may propagate across layers.
Estimating attacker capability, likelihood of success, or exploitability remains a separate technical task conducted outside the AIMS record (Section~\ref{sec:discussion}). 
Figure~\ref{fig:cedar-overview} summarizes the row-level processing procedure, while Table~\ref{tab:assessment-logic} distinguishes assessor-coded inputs from
engine-generated outputs.
\section{CEDAR-42001 Method}
\label{sec:method}

CEDAR-42001 converts coded ISO/IEC~42001 evidence into an integrated audit record that retains the conformity determination and supplements it with an audit-visible assurance diagnosis.
Stage~A establishes the conformity outcome.
Stage~B extends this outcome with a maturity assessment, identification of the binding constraint, architectural attribution, a risk-proportionate target, and an action recommendation.
The resulting records support decision-making at the strategic, operational, and tactical levels.

The method distinguishes between the \emph{assessment instrument} and the \emph{assessment engine}.
The assessment instrument specifies the audit questions, coding rubrics, architectural vocabulary, and decision rules. 
The assessment engine applies the conformity, maturity, target, and action rules to the coded records.
This separation allows the instrument and its rulebook to be reviewed or revised independently of the execution pipeline.

\subsection{Assessment Instrument and Evidence Coding}
\label{sec:method-instrument}

The assessment instrument comprises an ISO/IEC~42001 audit-question library, evidence-coding fields, maturity and layer-attribution rubrics, and a controls rulebook.
We developed the question library by decomposing Clauses~4--10 of the standard into atomic requirements.
Each independently assessable normative obligation was represented by a single audit question, resulting in a library of 159 questions.

The complete instrument includes the 159-question library, the 26-rule conformity hierarchy, the 17 IRDMO maturity gates, the target matrix, and the action templates.
It is released as a versioned executable artifact together with the Meridian fixture and Cruise coding used in Section~\ref{sec:evaluation} (Appendix~\ref{app:availability}).
The controls in Annex~A and the guidance in Annex~B inform the maturity and action rules but introduce no additional conformity requirements.

For each question, the assessor records the evidence source and accompanying note, the extent to which the requirement is satisfied, the implementation status and scope, and the severity context required for conformity classification.
The severity context comprises CPS impact, safety- or security-criticality, systemic scope, and repetition. 
The assessor also assigns five maturity scores and one primary architectural attribution. 
Secondary attributions may be retained when the evidence implicates additional layers.

When the available evidence is insufficient, the record is classified as \emph{Not assessed}, assigned no maturity level, and excluded from maturity aggregates.
In retrospective applications, this classification indicates that the selected evidence does not establish the relevant practice. 
It does not imply that undisclosed internal evidence was absent.

Once the coding process is complete, the assessment engine operates deterministically.
Identical coded inputs activate the same rules and generate the same findings, maturity levels, binding constraints, target gaps, and action recommendations.
This property ensures computational reproducibility for fixed inputs but does not imply agreement among independent assessors.

\begin{table}[t]
\centering
\caption{CEDAR-42001 row-level assessment logic. The assessor codes the
judgment-dependent inputs, and the assessment engine applies the corresponding
rule hierarchy.}
\label{tab:assessment-logic}
\scriptsize
\setlength{\tabcolsep}{3.0pt}
\renewcommand{\arraystretch}{1.08}
\begin{tabularx}{\textwidth}{@{}p{0.08\textwidth}
p{0.24\textwidth}
p{0.20\textwidth}
X@{}}
\toprule
\textbf{Step} &
\textbf{Assessor-coded inputs} &
\textbf{Recorded output} &
\textbf{Decision basis} \\
\midrule

Stage A &
Requirement satisfaction, implementation, scope, repetition, CPS impact,
systemicity, and safety/security criticality &
Conformity, OFI, Minor NC, Major NC, or Not assessed &
A priority-ordered hierarchy of 26 rules is applied. The first matching rule
determines the finding. \\

B1: Maturity &
Traceability, Operationalization, Monitoring, Improvement, and Cross-layer
Integration scores, each ranging from 0 to 4 &
Five-score profile, IRDMO level, and binding constraint &
A 17-rule gate maps the profile to Initial, Repeatable, Defined, Managed, or
Optimized. Unmet gates and unresolved nonconformities constrain progression. \\

B2: Attribution &
Primary and optional secondary layers assigned from the evidence note and
architectural rubric &
G, L1--L7, or Cross-layer &
The primary attribution identifies the location of the governed practice or
deficiency and supports portfolio-level aggregation. \\

B3: Target &
CPS impact and safety/security criticality &
Minimum target, optional stretch target, and target gap &
Higher-impact and safety- or security-critical practices are assigned more
demanding maturity targets. \\

B4: Action &
Finding, subclause, primary layer, binding constraint, target gap, and risk
context &
Corrective, improvement, assurance-strengthening, evidence-acquisition, or
sustainment action &
The finding determines the action class, while the remaining fields determine
the action content. \\

\bottomrule
\end{tabularx}
\end{table}

\subsection{Stage A: Conformity Determination}
\label{sec:method-stage-a}

Stage~A retains the conformity-assessment function of ISO/IEC~42001. A priority-ordered hierarchy of 26 rules maps each coded record to \emph{Conformity}, \emph{Opportunity for Improvement} (OFI), \emph{Minor Nonconformity}, \emph{Major Nonconformity}, or \emph{Not assessed}.

The coded inputs distinguish requirement satisfaction from severity context.
An unmet obligation is escalated when the evidence indicates systemic scope, high CPS impact, or safety- or security-critical consequences.
Repetition contributes to escalation only in the combinations specified by the rulebook.
An OFI is assigned when the obligation is satisfied, but the evidence supports a specific opportunity for improvement.
The first matching rule determines the finding and is recorded in the audit record. 
Stage~B retains this outcome and does not replace or modify it.

\subsection{Stage B: Audit-Visible Assurance Diagnosis}
\label{sec:method-stage-b}

Stage~B supplements the conformity outcome with four diagnostic components.

\paragraph{B1: Maturity profile and binding constraint.}
The assessor scores five dimensions on a scale from 0 to 4: \emph{Traceability}, \emph{Operationalization}, \emph{Monitoring}, \emph{Improvement}, and \emph{Cross-layer Integration}.  Collectively, these dimensions indicate whether a practice is adequately evidenced, consistently implemented, monitored, improved, and coordinated across the relevant governance and AI-CPS layers.
Appendix~\ref{app:rules} summarizes the scoring and decision rules; their complete executable definitions are included in the accompanying artifact.

A 17-rule gate maps the resulting profile to the IRDMO maturity ladder: \emph{Initial}, \emph{Repeatable}, \emph{Defined}, \emph{Managed}, and \emph{Optimized}.
The Managed level requires sufficient Monitoring and Improvement scores, whereas the Optimized level additionally requires Cross-layer Integration.
Unresolved nonconformities impose an upper limit on the level that may be attained. The binding constraint is the score or gate
that prevents progression to the next applicable level.

These five levels follow the SW-CMM lineage~\cite{ISACA-CMMI}. 
CEDAR-42001 does not claim the maturity ladder itself as a novel contribution.
Its contribution lies in the gate logic that relates each maturity level to the five evidence dimensions and to the ceiling imposed by unresolved nonconformities.

\paragraph{B2: Governance and AI-CPS layer attribution.}
Each record receives one primary attribution from the governance layer (G), the seven AI-CPS layers (L1--L7), or the Cross-layer reporting category defined in Section~\ref{sec:assurance-scope}.
The assessor assigns the attribution based on the evidence note and the layer rubric. 
The assessment engine then validates the assignment and uses it for aggregation and action generation.

The attribution identifies the location of the governed practice or deficiency, which may differ from the location at which the resulting harm becomes observable.
A consequence at L1 may therefore be attributed primarily to perception at L2, decision-making at L5, or supervision at L7. 
Secondary attributions preserve additional relationships between layers. 
Portfolio-level statistics, however, use a single primary category to avoid double-counting.

\paragraph{B3: Risk-proportionate target maturity.}
The assessment engine assigns a minimum target and, where applicable, a stretch target based on CPS impact and safety- or security-criticality. 
Practices with higher potential impact or greater criticality require stronger institutionalization, monitoring, improvement, and cross-layer integration.
Each record specifies the assessed maturity level, the applicable target, and the resulting target gap.
Achievement of the target does not, by itself, establish that the underlying system is safe or secure.

\paragraph{B4: Action recommendation.}
The Stage~A finding determines the action class.
Nonconformities generate corrective actions, while OFIs generate improvement actions.
Conforming records that remain below their target generate assurance-strengthening recommendations, and Not-assessed records generate evidence-acquisition actions.
Conforming records that meet the applicable target may receive a sustainment note.

The action content is determined by the ISO/IEC~42001 subclause, the primary layer, the binding constraint, and the risk context.
Specific rules take precedence over general rules.
When no specific rule applies, the assessment engine uses a traceable template based on the relevant subclause and finding. 
The term Corrective and Preventive Action (CAPA) is reserved for actions associated with nonconformities.

\subsection{Integrated Audit Record and Decision Products}
\label{sec:method-rows}

The integrated audit record retains the question identifier, ISO/IEC~42001 source, evidence note, Stage~A finding, and triggering rule. 
It supplements these fields with five maturity scores, IRDMO level, binding constraint, architectural attribution, target maturity, target gap, and action recommendations.

Table~\ref{tab:integrated-rows} presents two records from the Cruise
application.
Both records are classified as Major Nonconformities.
Q87 concerns governance-level communication and disclosure, whereas Q113 concerns decision-layer post-collision control behavior. 
Although the two records have the same conformity severity, they differ in their architectural attribution, binding constraint, and recommended action.

\begin{table}[t]
\centering
\caption{Integrated CEDAR-42001 records for two selected Cruise questions.
Scores are reported in the following order:
Traceability/Operationalization/Monitoring/Improvement/Cross-layer
Integration.}
\label{tab:integrated-rows}
\scriptsize
\setlength{\tabcolsep}{2.7pt}
\renewcommand{\arraystretch}{1.08}
\begin{tabularx}{\textwidth}{@{}p{0.08\textwidth}
p{0.11\textwidth}
p{0.08\textwidth}
p{0.25\textwidth}
p{0.17\textwidth}
X@{}}
\toprule
\textbf{Q.} &
\textbf{Stage A} &
\textbf{Layer} &
\textbf{Maturity diagnosis} &
\textbf{Target and gap} &
\textbf{Action recommendation} \\
\midrule

Q87
(7.4) &
Major NC &
G &
$2/1/0/0/0 \rightarrow$ Initial.
Monitoring is the immediate binding constraint. &
Minimum: Managed;
stretch: Optimized;
gap: Yes &
Establish controls to ensure disclosure completeness, independent review, and
monitoring of material omissions. \\

Q113
(8.1) &
Major NC &
L5 &
$2/1/1/0/1 \rightarrow$ Initial.
Improvement is the binding constraint. &
Minimum: Managed;
stretch: Optimized;
gap: Yes &
Validate post-collision control logic and define monitored fail-safe or hold behavior for unanticipated states. \\

\bottomrule
\end{tabularx}
\end{table}

The integrated records are aggregated into three complementary decision views.
The \emph{strategic} view summarizes conformity outcomes, concentrations of risk, and recurring capability constraints. 
The \emph{operational} view groups findings by architectural layer, maturity gap, and binding constraint.
The \emph{tactical} view serves as the action register.
Because all three views are derived from the same underlying records, each aggregate result remains traceable to the corresponding obligation, evidence note, coded judgment, and triggered rule.
\section{Application and Coverage Study}
\label{sec:evaluation}

We demonstrate two bounded properties of CEDAR-42001.
First, Section~\ref{sec:meridian-case} examines whether the complete method operates end-to-end on a fully coded, document-level synthetic AIMS evidence document.
Second, Section~\ref{sec:eval-coverage} examines whether Stage~B produces diagnostic information that is absent from a conformity-only ledger. 
Section~\ref{sec:cruise-application} then illustrates retrospective applicability using the documented 2023 Cruise robotaxi incident. These analyses demonstrate pipeline coverage and representational value; they do not constitute inferential validation.

\subsection{Evidence-Backed Pipeline Demonstration: Meridian Robotics}
\label{sec:meridian-case}

\paragraph{Synthetic AIMS evidence corpus.}
We constructed Meridian Robotics as a controlled synthetic case of an organization that develops, deploys, and operates autonomous mobile robots.
Rather than assigning findings directly, we developed a structured AIMS evidence corpus for organizational policy. 
The policy includes records on organizational context and scope, an AI asset register, risk management and treatment, a Statement of Applicability, operational and change-management procedures, competence, monitoring, management review, nonconformity and corrective action, and selected supplier and internal communications.
The complete corpus and evidence-to-question mapping are provided in the accompanying artifact.

The corpus contains heterogeneous and, in some cases, conflicting evidence.
Meridian exhibits mature functional-safety practices, including an independent emergency-stop chain, validated stopping distances, controlled software deployment, safety-performance monitoring, and closed-loop treatment of safety-related near misses. 
Its AI-specific governance is less mature.
Documented weaknesses include no in-service model-performance monitoring, incomplete assurance of a supplier-provided perception model, undefined human-oversight and teleoperation procedures, and exclusion of the fleet communication network from the AIMS scope despite its role in transmitting command and stop messages. 
The corpus also contains inconsistencies between documented claims and operational evidence, including controls recorded as implemented in the Statement of Applicability without corresponding operational evidence. 
The case, therefore, tests distinctions among conformity outcomes, maturity profiles, architectural layers, risk targets, and action classes rather than assuming that practices are uniformly weak or mature.

\paragraph{Assessment procedure.}
The instrument was applied row by row.
For each question, the assessor identified the relevant source and recorded an evidence note, requirement-satisfaction judgment, implementation and scope judgments, severity context, scores for five maturity dimensions, and a primary architectural attribution. 
These fields are evidence-grounded assessor judgments, not engine-generated results.
The engine then applied the conformity hierarchy, IRDMO gates, target rules, and action hierarchy to generate the Stage~A finding, maturity level, binding constraint, risk-proportionate target, target gap, and recommended action.

The evaluation distinguishes among documentary evidence, assessor-coded interpretations, and deterministic engine outputs derived from the coded rows.
Findings and Stage~B outputs were generated from coded evidence rather than entered as predetermined results.
Meridian is therefore an evidence-backed synthetic audit case for fully evaluating the pipeline and representational coverage, not an estimate of findings or maturity levels in operational organizations.

\subsection{Pipeline Coverage on a Synthetic AMR-Fleet AIMS}
\label{sec:eval-coverage}

\paragraph{Pipeline and representational coverage.}
All 159 questions were assessed against the Meridian corpus. 
Stage~A produced 143 Conformity findings, eight Minor Nonconformities, seven Major Nonconformities, and one OFI.
Nine of the 26 conformity rules were activated, and no row required an unclassified default. 
The case, therefore, covers every processing stage, but not every conformity-rule branch.

The rows covered the governance layer and all seven AI-CPS architectural layers. 
Governance accounted for 127 rows. 
The seven Major Nonconformities were distributed across governance (G), sensing and perception (L2), middleware and communication (L4), and AI decision-making and control (L5).
Meridian thus demonstrates the translation of management-system obligations into architectural locations rather than concentrating all weaknesses in the governance layer.

\begin{figure}[t]
\centering
\includegraphics[width=0.74\textwidth]
{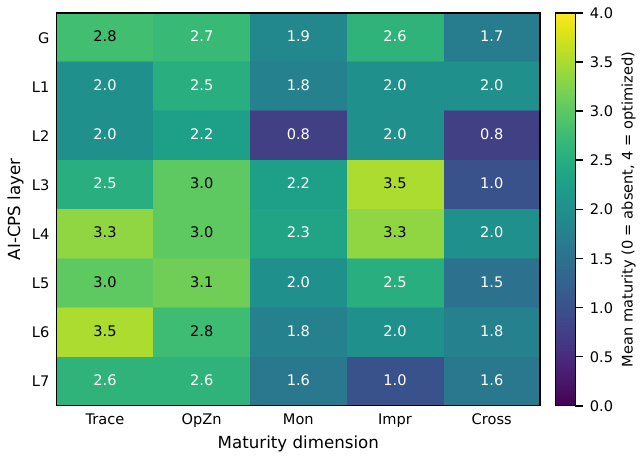}
\caption{Mean maturity scores by governance layer and AI-CPS architectural layer in the Meridian fixture. Monitoring and Cross-layer Integration are the lowest-scoring dimensions overall, while layer-specific profiles reveal distinct binding constraints that are not visible from conformity counts alone.}
\label{fig:meridian-maturity}
\end{figure}

\paragraph{Maturity and binding constraints.}
Figure~\ref{fig:meridian-maturity} presents maturity by layer and dimension. 
Traceability and Operationalization have the highest organization-wide means, at 2.78 and 2.72, respectively, whereas Monitoring and Cross-layer Integration have the lowest, at 1.86 and 1.65, respectively. 
These weaker dimensions repeatedly prevent progression to the Managed or Optimized levels. 
The layer-level view also reveals patterns not apparent from the count-based findings: L2 has mean scores of 0.8 for both Monitoring and Cross-layer Integration, whereas L7 has a mean Improvement score of 1.0.
Conformity severity and capability maturity, therefore, provide distinct perspectives on assurance needs.

\paragraph{Conformity-assurance gap.}
The baseline assurance band is the mean of Monitoring, Improvement, and Cross-layer Integration.
Means below 1.5 are classified as Low, from 1.5 to below 2.5 as Medium, and at least 2.5 as High. 
This is a CEDAR-42001 analytical rule, not an ISO/IEC~42001 requirement.

\begin{figure}[t]
    \centering
    \includegraphics[width=0.72\linewidth]{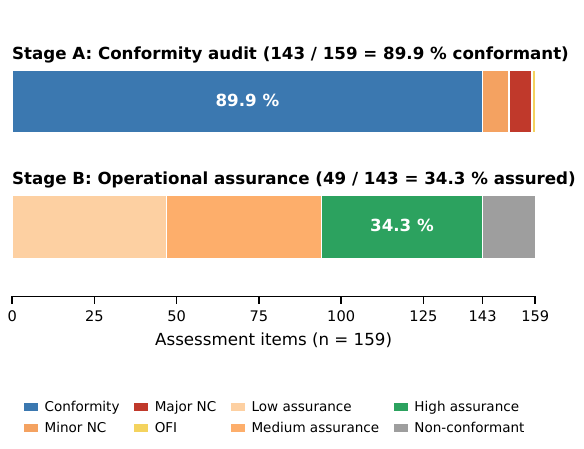}
    \caption{Stage~A conformity and Stage~B assurance outcomes in the Meridian fixture. Of 159 assessed rows, 143 were conforming; 49 of those conforming rows reached the baseline high-assurance category.}
    \label{fig:audit-assurance-gap}
\end{figure}

Out of the 159 rows, 143 are conforming (89.9\%), but only 49 of those 143 rows (34.3\%) reach the baseline High-assurance category (Figure~\ref{fig:audit-assurance-gap}).
This separation describes the evidence corpus and its coding, not assurance prevalence in operational organizations.

Conformity is calculated over all 159 rows, whereas assurance categories are calculated over the 143 conforming rows. 
A conforming row below High assurance indicates insufficient evidence of Monitoring, Improvement, or Cross-layer Integration, not necessarily an ineffective control.
Governance accounts for 127 rows, while the remaining 32 layer-attributed rows provide the primary basis for the architectural diagnosis, including Major Nonconformities across G, L2, L4, and L5.

\paragraph{Threshold and aggregation sensitivity.}
Because the baseline category depends on author-defined dimensions, aggregation, and thresholds, we recalculated the High-assurance proportion under four alternative operationalizations.
The proportion ranged from 22.4\% to 46.2\%, and most conforming rows remained below High assurance under every specification.
Appendix~\ref{app:sensitivity} reports the corresponding criteria, counts, and percentages.
Only the threshold and aggregation rules were varied; the evidence coding, maturity scores, conformity findings, target rules, and architectural attributions remained fixed.

\paragraph{Risk targets and pipeline conclusion.}
Meridian contains seven High-impact findings, five of which are also safety- or security-critical and therefore receive the strongest applicable minimum and stretch targets. 
Across the 159 rows, the case exercises every CEDAR-42001 processing stage, all finding outcomes instantiated in Meridian, the complete IRDMO ladder, the governance layer and layers L1-L7, target-gap generation, and all applicable action classes. 
Because it activates nine of the 26 conformity rules, the case demonstrates full pipeline coverage rather than exhaustive rule-branch coverage.

These results support the bounded claim that CEDAR-42001 operates end-to-end on a fully coded, document-level synthetic AIMS evidence corpus and produces the intended integrated rows and aggregate decision artifacts.
They do not establish the prevalence of comparable findings in operational organizations, agreement among independent assessors, or the effectiveness of the recommendations.

\subsection{Retrospective Application to the Cruise Robotaxi Incident}
\label{sec:cruise-application}

This subsection evaluates whether CEDAR-42001 can represent documented governance, perception, decision-making, and oversight concerns associated with a single incident.
It does not assess diagnostic or predictive performance.

We apply CEDAR-42001 retrospectively to the public record of the October 2023 Cruise robotaxi incident~\cite{quinnemanuel2024,koopman2024,cpuc2023osc,dotoig2024,nhtsa2023recall,nhtsa2024consent}. 
Cruise is treated as an analytical case rather than a conformant AIMS.
Because the same public sources inform both coding and interpretation, the case is an application of the framework rather than an independent validation.

Of the 159 questions, 51 could be assessed, and 108 were marked \emph{Not assessed}. 
Fifty assessable rows received nonconformity findings, and one received a Conformity finding.
This distribution is expected from failure-centric public records and demonstrates coverage of documented concerns rather than the organization's overall posture.

The concerns cluster in four locations. 
Governance (G) findings address communication completeness, escalation, management review, and corrective response.
Sensing and perception (L2) findings concern failure to maintain an adequate representation of the pedestrian beneath the vehicle.
Decision-making and control (L5) findings concern execution of the post-collision pullover maneuver rather than maintenance of a stationary state. 
Human supervision (L7) findings concern whether remote assistance had sufficient time, authority, and interlock capability to prevent further movement.

The physical consequence occurred at L1, whereas the governing weaknesses were mainly attributed to G, L2, L5, and L7. 
Thus, the layer in which harm manifests need not be the layer in which assurance is deficient. 
These attributions also yield differentiated actions: governance controls address disclosure, accountability, review, and omission monitoring; L2 actions address perception validation and expanded scenario coverage; L5 actions address control-logic validation and stationary or fail-safe behavior; and L7 actions address intervention timing, hold states, confirmation, and interlocks.

The Cruise application, therefore, shows that CEDAR-42001 can encompass distinct governance, perception, decision-making, and oversight concerns within a single documented incident.
\section{Discussion and Limitations}
\label{sec:discussion}

\paragraph{Conformity and assurance provide distinct signals.} CEDAR-42001 distinguishes compliance with a management-system obligation from the assurance supported by the maturity of the associated practice. In the Meridian fixture, 89.9\% of rows received a Stage~A Conformity finding, yet only 34.3\% of conforming rows reached the baseline High-assurance category. The proportion remained between 22.4\% and 46.2\% under the alternative threshold and aggregation specifications evaluated in Section~\ref{sec:eval-coverage}; the corresponding specification-level results are reported in Appendix~\ref{app:sensitivity}. This conformity-assurance gap reflects limited evidence of monitoring, improvement, and cross-layer integration rather than evidence that the underlying controls were technically ineffective. Because Meridian is synthetic, these proportions characterize the fixture and its coding rather than the prevalence of assurance weaknesses among organizations implementing ISO/IEC~42001.

\paragraph{Architecture shapes the operational interpretation of findings.}
Clause-level grouping identifies the relevant management-system obligation but not necessarily the architectural location of the weakness. The maturity profile in Figure~\ref{fig:meridian-maturity} and the distribution of Major Nonconformities across G, L2, L4, and L5 provide a complementary operational view. Findings with the same conformity severity may require different responses depending on whether they concern governance, perception, communication, decision-making, or oversight. The integrated audit row preserves this distinction without changing the original conformity decision.

CEDAR-42001 is therefore a governance-to-assurance handoff rather than a technical CPS security assessment. It does not evaluate attack surfaces, software vulnerabilities, controller behavior, process invariants, runtime telemetry, or adversarial robustness. Instead, it identifies where governance evidence indicates a need for deeper technical assurance, organizational improvement, or remediation. Penetration testing, controller validation, runtime monitoring, and safety analysis require evidence beyond the AIMS audit record.

\paragraph{Decision support is informational rather than causal.}
As defined in Section~\ref{sec:method-rows} and demonstrated in Section~\ref{sec:evaluation}, Stage~B supplements the Stage~A conformity ledger with strategic information on risk concentrations and recurring capability constraints, operational information on architectural attribution and target gaps, and tactical action recommendations. The evaluation shows that these outputs are not available from the conformity result alone. It does not, however, establish that decision-makers will allocate resources more effectively, implement the recommendations correctly, or prevent future incidents.

\paragraph{Interpretation of the Cruise application.}
The Cruise case demonstrates retrospective applicability rather than source-independent validation. Because the same public record informs both coding and interpretation, the correspondence with documented governance, perception, decision-making, and oversight concerns is partly expected. Public incident reports also emphasize failures and rarely document satisfactory routine practices. Thus, the 51 assessable and 108 Not-assessed rows characterize only the audit-visible posture supported by the selected evidence, not Cruise's complete internal practices.

\paragraph{Construct, instrument, and reliability limitations.}
The assurance construct, maturity gates, target rules, and action templates were defined by the author. Although sensitivity analysis reduces reliance on a single specification, it neither identifies a uniquely correct threshold nor externally validates the construct. The 159-question instrument was derived by one analyst through atomic decomposition of ISO/IEC~42001, and alternative decompositions remain possible.

For a fixed coded input vector, the engine operates deterministically and is computationally reproducible. This does not establish inter-assessor reliability, as assessors may differ in judgments of evidence sufficiency, contextual severity, maturity, or architectural attribution. The Cruise classifications have not been shown to be uniquely correct, reproducible prospectively, or generalizable across AI-CPS domains. Future evaluations should examine operational AIMSs, involve multiple assessors with ISO/IEC~42001 and CPS-security expertise, and compare results with independent technical-assurance findings.
\section{Conclusion}
\label{sec:conclusion}

CEDAR-42001 translates ISO/IEC~42001 conformity evidence into an architecture-aware, audit-visible assurance posture for AI-enabled cyber-physical systems. Stage~A preserves the original conformity determination, while Stage~B adds architectural attribution, maturity, and binding-constraint diagnosis, risk-proportionate assurance targets, and actionable recommendations.

The Meridian fixture demonstrates end-to-end applicability and shows that conformity and audit-visible assurance represent distinct dimensions of system governance. Although 89.9\% of its rows were conforming, only 34.3\% reached the baseline High-assurance category, with proportions ranging from 22.4\% to 46.2\% across alternative operationalizations. These results characterize the Meridian fixture rather than the prevalence of such gaps in operational organizations. The retrospective Cruise application further illustrates the framework's ability to differentiate governance, perception, decision-making, and oversight concerns within a single case.

CEDAR-42001 does not replace technical cybersecurity or safety assessment; it identifies where audit evidence indicates a need for deeper technical assurance, organizational improvement, or targeted remediation.

\bibliographystyle{splncs04}
\bibliography{references}
\appendix

\section{Artifact and Data Availability}
\label{app:availability}

The CEDAR-42001 instrument, rulebook, execution engine, Meridian fixture, and Cruise coding used in Section~\ref{sec:evaluation} are available at \href{https://doi.org/10.5281/zenodo.20722090}{repository} under version 1.0. The artifact contains the 159-question library, the 26-rule conformity hierarchy, the 17 IRDMO maturity gates, the target matrix, the action templates, the coded row-level inputs, and the generated outputs.
It therefore supports re-execution of the reported analyses for the published coding, but not independent replication of the assessor judgments.

\section{Maturity and Decision-Rule Summary}
\label{app:rules}

Table~\ref{tab:maturity-dimensions} summarizes the five Stage~B maturity dimensions.
Each dimension is scored from 0 to 4: 0 indicates no supporting
evidence; 1, an informal or ad hoc practice; 2, a documented or repeatable but incomplete practice; 3, consistent and evidenced operation; and 4, demonstrated continual improvement or cross-layer integration.

\begin{table}[t]
\centering
\caption{Stage~B maturity dimensions.}
\label{tab:maturity-dimensions}
\small
\begin{tabular}{p{0.22\linewidth}p{0.70\linewidth}}
\hline
\textbf{Dimension} & \textbf{Evidence represented} \\
\hline
Traceability &
Responsibilities, decisions, evidence, and changes are linked to the governed
obligation. \\
Operationalization &
The documented practice is implemented consistently within its intended scope. \\
Monitoring &
Indicators, reviews, or observations provide evidence of operation and
effectiveness. \\
Improvement &
Identified deficiencies lead to controlled change and organizational learning. \\
Cross-layer Integration &
The practice is coordinated across the relevant governance and AI-CPS layers. \\
\hline
\end{tabular}
\end{table}

The 17-rule gate maps the resulting profile to one of the following maturity levels: Initial, Repeatable, Defined, Managed, or Optimized.
Managed requires sufficient Monitoring and Improvement, while Optimized additionally requires Cross-layer Integration.
Unresolved nonconformities constrain the maximum attainable level. 
The binding constraint is the unmet score or gate that prevents progression.

Targets are assigned from CPS impact and safety- or security-criticality.
Higher-impact or critical practices receive stronger minimum and, where applicable, stretch targets. 
The Stage~A finding determines the action class:
Major and Minor Nonconformities generate corrective actions; OFIs generate improvement actions; conforming rows below the target generate assurance-strengthening actions; conforming rows meeting target generate sustainment actions; and Not-assessed rows generate evidence-acquisition actions.
The subclause, primary layer, binding constraint, and risk context
determine the action content.
Complete gate definitions, target rules, rule precedence, and templates are provided in the artifact described in Appendix~\ref{app:availability}.

\section{Conformity--Assurance Sensitivity Analysis}
\label{app:sensitivity}

The baseline High-assurance category uses the mean of the Monitoring, Improvement, and Cross-layer Integration categories. Because this operationalization is author-defined, we recalculated the classification of the 143 conforming Meridian rows under four alternatives. 
Only the aggregation and threshold criteria were varied; evidence coding, conformity findings, maturity scores, architectural attributions, and target rules remained fixed.

\begin{table}[t]
\centering
\caption{Sensitivity of the Meridian conformity--assurance gap. Percentages are
calculated over the 143 rows receiving a Stage~A Conformity finding.}
\label{tab:assurance-sensitivity}
\small
\begin{tabular}{p{0.18\linewidth}p{0.47\linewidth}p{0.14\linewidth}p{0.14\linewidth}}
\hline
\textbf{Specification} & \textbf{High-assurance criterion} &
\textbf{High} & \textbf{Below High} \\
\hline
Baseline &
Mean of Monitoring, Improvement, and Cross-layer Integration $\geq 2.50$ &
49 (34.3\%) & 94 (65.7\%) \\

Lower threshold &
Three-dimension mean $\geq 2.25$ &
63 (44.1\%) & 80 (55.9\%) \\

Higher threshold &
Three-dimension mean $\geq 2.75$ &
32 (22.4\%) & 111 (77.6\%) \\

Floor-constrained &
Three-dimension mean $\geq 2.50$, with every dimension $\geq 2$ &
43 (30.1\%) & 100 (69.9\%) \\

Five-dimension mean &
Mean of all five maturity dimensions $\geq 2.50$ &
66 (46.2\%) & 77 (53.8\%) \\
\hline
\end{tabular}
\end{table}

Across the tested specifications, the High-assurance proportion ranges from 22.4\% to 46.2\%; consequently, most conforming rows remain below the High-assurance category under every operationalization.
These results characterize the Meridian fixture and its coding rather than the prevalence of assurance weaknesses in operational organizations.

\end{document}